\documentclass[twocolumn,preprintnumbers]{revtex4}
\usepackage{graphics}
\usepackage{epsfig}
\usepackage{amsmath}
\usepackage{amssymb}

\newcommand{\beq}{\begin{equation}}
\newcommand{\eeq}{\end{equation}}
\newcommand{\bea}{\begin{eqnarray}}
\newcommand{\eea}{\end{eqnarray}}
\newcommand{\cir}{{\buildrel \circ \over =}}

\renewcommand{\hat}{{}}

\begin{document}

\title{Background-Independent Gravitational Waves.}

\author{Juri Agresti}
\email{agresti@fi.infn.it}
\affiliation{
Dipartimento di Fisica,  Universit\`a di Firenze,
via Sansone 1, 50019 Sesto Fiorentino, Italy}

\author{Roberto De Pietri}
\email{depietri@pr.infn.it}
\affiliation{
Dipartimento di Fisica, Universit\`a di Parma e INFN Parma,
Parco delle Scienze 7/A, 43100 Parma, Italy}
\author{Luca Lusanna}
\email{lusanna@fi.infn.it}
\affiliation{
Sezione INFN di Firenze, via Sansone 1,
50019 Sesto Fiorentino, Italy}

\author{Luca Martucci}
\email{luca.martucci@mi.infn.it}
\affiliation{
Dipartimento di Fisica, Universit\`a di Milano
e INFN Sezione di Milano, via G.Celoria  16, 20133 Milano}
\date{\today}
\begin{abstract}
A Hamiltonian linearization of the rest-frame instant form of
tetrad gravity, where the Hamiltonian is the weak ADM energy
${\hat E}_{ADM}$, in a completely fixed (non harmonic)
3-orthogonal Hamiltonian gauge is defined. For the first time this
allows to find an explicit solution of all the Hamiltonian
constraints and an associated linearized solution of Einstein's
equations. It corresponds to background-independent gravitational
waves in a well defined post-Minkowskian Christodoulou-Klainermann
space-time.
\end{abstract}

\maketitle

ADM canonical gravity has always been important for the Cauchy
problem and for the interpretation of general relativity, but till
now it has not produced new exact solutions of Einstein's
equations due to the difficulty in solving its first class
constraints. Instead recently there have been the first attempts
to use it for numerical simulations of stars and of the coalescence
of binary systems of neutron stars and/or black holes.
In this letter we will show that it is possible to find an approximate
solution of all the constraints and then of vacuum Einstein equations by
defining a Hamiltonian linearization of ADM tetrad gravity in a completely
fixed Hamiltonian gauge, corresponding to a system of non-harmonic
4-coordinates. This approximate solution corresponds to a {\it
background-independent gravitational wave} and defines a
post-Minkowskian space-time, linearization of a Christodoulou -
Klainermann space-time \cite{1}. Here only the general method and
the main results will be stated, since all the detailed
calculations are contained in Ref.\cite{2}.

As shown in Ref.\cite{3}, it is possible to define a {\it
rest-frame instant form} of ADM metric gravity, without or with
matter, for globally hyperbolic, topologically trivial,
asymptotically flat at spatial infinity space-times, in such a way
that in the limit of vanishing Newton constant the rest-frame
instant form of parametrized Minkowski theories \cite{4} is
recovered. This is possible if the allowed 3+1 splittings of the
space-time are associated with foliations whose leaves are
space-like hyper-surfaces $\Sigma^{(WSW)}_{\tau}$ [named
Wigner-Sen-Witten (WSW) surfaces \cite{2}; $\tau$ is a
mathematical time labeling the leaves; $\vec \sigma$ are
curvilinear coordinates on $\Sigma_{\tau}$] tending, in a
direction-independent way, to Minkowski hyper-planes orthogonal to
the ADM 4-momentum at spatial infinity \cite{5}. The
{\it SPI group} of asymptotic symmetries at spatial infinity is reduced
to the Poincar\`e group with the strong ADM Poincar\`e charges as
generators. The 4-metric is assumed to have a well defined
direction-independent limit at spatial infinity. It can be shown
\cite{3} that in the rest-frame instant form of gravity the Dirac
Hamiltonian, after the addition of a suitable surface term and a
suitable splitting of the lapse and shift functions in asymptotic
and bulk parts, is weakly equal to the {\it weak ADM energy} (see
also Ref.\cite{6}), which is an integral over 3-space, weakly
equal to the strong ADM energy (surface form). In Ref.\cite{7},
after the introduction of a new parametrization of tetrad gravity
\cite{8}, its rest-frame instant form  is introduced. In this form
of dynamics we have the extra constraints ${\vec P}_{ADM} \approx
0$ ,
defining the {\it rest frame of the universe} like in
Christodoulou - Klainermann space-times.

The canonical variables of this re-formulation of tetrad gravity
are 3 boost parameters $\varphi_{(a)}(\tau ,\vec \sigma )$, 3
angles $\alpha_{(a)}(\tau ,\vec \sigma )$, the bulk parts $n(\tau
,\vec \sigma )$, $n_r(\tau ,\vec \sigma )$ of the lapse and shift
functions, cotriads ${}^3e_{(a)r}(\tau ,\vec \sigma )$ over
$\Sigma^{(WSW)}_{\tau}$ and the 16 conjugate momenta. There are 14
first class constraints: 7 are given by the vanishing of the
boost, lapse and shift momenta; 6 are the generators of rotations
and passive 3-diffeomorphisms and the last one is the
super-hamiltonian constraint. This last constraint is the
generator of the Hamiltonian gauge transformations which modify an
allowed 3+1 splitting of space-time into another allowed one: as a
consequence the physical results are independent from the choice
of the 3+1 splitting. It can be shown \cite{3} that the
super-hamiltonian constraint is an equation for the determination
of the conformal factor $\phi (\tau ,\vec \sigma ) = det(\,
{}^3g(\tau ,\vec \sigma )\, )^{1/12}$ of the 3-metric on
$\Sigma_{\tau}$, namely it has to be interpreted as the
Lichnerowicz equation.

By means of a point Shanmugadhasan canonical transformation
\cite{9} 13 constraints (with the exception of the
super-hamiltonian one) become momenta of the new canonical basis
(Abelianization of the first class constraints). The conjugate
variables are 13 Abelianized gauge variables (describing
generalized inertial effects). The canonical basis is
completed with i) the conformal factor $\phi(\tau ,\vec \sigma )$
(to be determined by the Lichnerowicz equation) and
$\pi_{\phi}(\tau ,\vec \sigma )$; ii) two (non-tensorial)
conjugate pairs $r_{\bar a}(\tau ,\vec \sigma )$,
$\pi_{\bar a}(\tau ,\vec \sigma )$, $\bar a=1,2$, of Dirac
observables (DO; the deterministically predictable degrees of
freedom of the gravitational field, describing generalized tidal
effects). See Refs.\cite{3,7,10} for the interpretational
problems.

By adding 14 suitable gauge fixing constraints (one of them must
be $\pi_{\phi}(\tau ,\vec \sigma ) \approx 0$), we get a
completely fixed 3-orthogonal Hamiltonian gauge in which the
3-metric is diagonal. It corresponds, on the solutions of
Einstein's equations, to a unique non-harmonic 4-coordinate system
\cite{2}. In this radiation gauge all  4-tensors are expressed
only in terms of the two pairs of conjugate DO $r_{\bar a}(\tau
,\vec \sigma )$, $\pi_{\bar a}(\tau ,\vec \sigma )$. In particular
we get ${}^3g_{rs} = \delta_{rs}\, \phi^4\, exp( {2\over
{\sqrt{3}}}\, \sum_{\bar a}\, \gamma_{\bar ar}\, r_{\bar a})$,
where the $\gamma_{\bar ar}$'s are suitable constants. The DO
$r_{\bar a}(\tau ,\vec \sigma )$ {\it replace the two
polarizations of the harmonic TT gauge}. Let us remark that in
absence of a background we do not have a theory of a spin 2 wave
propagating over Minkowski space-time but a genuine Einstein
space-time.

It can be shown \cite{2} that in the new canonical basis the
solution of the 6 rotation and super-momentum constraints reduces
to solve a set of quasi-linear partial differential equations of
elliptic type. Instead the Lichnerowicz equation becomes a
in-tractable integro-differential equation for $\phi (\tau ,\vec
\sigma )$. Also the bulk lapse and shift functions are determined
by integral equations.

Then, without never introducing a background 4-metric, we make a
{\it weak field approximation} and a Hamiltonian linearization
by systematically discarding terms of order $O(r^2_{\bar a})$ in
the Lichnerowicz equation and $O(r^3_{\bar a})$ in the weak ADM
energy.

A) We assume $| r_{\bar a}(\tau ,\vec \sigma ) | << 1$ on each WSW
hyper-surface and $|\partial_u r_{\bar a}(\tau ,\vec \sigma )|
\sim {1\over L} O(r_{\bar a})$, $|\partial_u\partial_v r_{\bar
a}(\tau ,\vec \sigma )| \sim {1\over {L^2}} O(r_{\bar a})$, where
$L$ is a {\it big enough characteristic length interpretable as
the reduced wavelength $\lambda / 2\pi$ of the resulting
gravitational waves}. Since the conjugate momenta $\pi_{\bar
a}(\tau ,\vec \sigma )$ have the dimensions of ${{action}\over
{L^3}}$, i.e. of ${k\over L}$ with $k={{c^3}\over {16\pi G}}$, we
assume $|\pi_{\bar a}(\tau ,\vec \sigma )| \sim {k\over L}
O(r_{\bar a})$, $|\partial_u \pi_{\bar a}(\tau ,\vec \sigma )|
\sim {k\over {L^2}} O(r_{\bar a})$, $|\partial_u\partial_v
\pi_{\bar a}(\tau ,\vec \sigma )| \sim {k\over {L^3}} O(r_{\bar
a})$. Therefore, $r_{\bar a}(\tau ,\vec \sigma )$ and $\pi_{\bar
a}(\tau ,\vec \sigma )$ are {\it slowly varying over the length L}
(for $r_{\bar a}, \pi_{\bar a}\, \rightarrow \, 0$ we get the void
space-times of Ref.\cite{3}). It can be shown \cite{2} that the
Riemann tensor of our space-time is of order ${1\over {L\, k}}\,
O(\pi_{\bar a}) = {1\over {L^2}}\, O(r_{\bar a}) \approx {\cal
R}^{-2}$, where ${\cal R}$ is the mean radius of curvature.
Therefore the {\it requirements of the weak field approximation
are satisfied}: i) ${\cal A} = O(r_{\bar a})$, if ${\cal A}$ is
the amplitude of the gravitational wave; ii) ${L\over {{\cal R}}}
= O(r_{\bar a})$, namely ${{\lambda}\over {2\pi}} << {\cal R}$.

B) We also assume $q(\tau ,\vec \sigma ) = 2\, ln\, \phi (\tau
,\vec \sigma ) \sim O(r_{\bar a})$, $\partial_u q(\tau ,\vec
\sigma ) \sim {1\over L} O(r_{\bar a})$,
$\partial_u\partial_vq(\tau ,\vec \sigma ) \sim {1\over {L^2}}
O(r_{\bar a})$, so that we get $\phi (\tau ,\vec \sigma ) =
e^{q(\tau ,\vec \sigma )/2} \approx 1+ {1\over 2} q(\tau ,\vec
\sigma ) +O(r^2_{\bar a})$ for the conformal factor. The
Lichnerowicz equation becomes the linear partial differential
equation  $\triangle q(\tau ,\vec \sigma ) = {1\over {2\sqrt{3}}}
\sum_{u\bar a} \gamma_{\bar au} \partial^2_ur_{\bar a}(\tau ,\vec
\sigma ) +{1\over {L^2}} O(r_{\bar a}^2)$ for $q(\tau ,\vec \sigma
)$,  whose solution vanishing at spatial infinity  is
\begin{equation}
 q(\tau ,\vec \sigma ) =
 {-1\over {2\sqrt{3}}} \sum_{u\bar a}
\gamma_{\bar au} \int d^3\sigma_1 {{\partial^2_{1u}r_{\bar a}(\tau
,{\vec \sigma}_1)} \over {4\pi |\vec \sigma -{\vec \sigma}_1|}}
+O(r^2_{\bar a}).
 \label{1}
\end{equation}
Then we are able to solve the quasi-linear partial differential
equations equivalent to the 6 rotation and super-momentum
constraints. Their solution implies that the cotriad momenta have
the following expression in terms of DO's
\begin{eqnarray} \label{2}
&&
\!\!\!\!{}^3\pi^r_{(a)}(\tau ,\vec \sigma ) =\sqrt{3}
 \sum_{\bar a}\gamma_{\bar ar} \delta_{(a)}^r \pi_{\bar a}(\tau
 ,\vec \sigma )+
\\
&&\,  + {{\sqrt{3}}\over 2}[1-\delta^r_{(a)}] \sum_{\bar au} \gamma_{\bar au}
  [1-2(\delta_{ru}+\delta_{au})] \,\cdot
\nonumber \\
&&\,  \cdot\, {{\partial^2}\over {(\partial \sigma^u)^2}}
  \int^{\infty}_{\sigma^r} d\sigma_1^r
  \int^{\infty}_{\sigma^a} d\sigma^a_1\, \pi_{\bar a}(\tau ,\sigma_1^r
  \sigma_1^a\sigma^{k\not= r,a}) + O(r^2_{\bar a}),
\nonumber
\end{eqnarray}
\noindent with the restriction $\int_{-\infty}^{+\infty}
d\sigma^r\, \pi_{\bar a}(\tau ,\vec \sigma ) = 0$, $r=1,2,3$.
Hamilton equations are compatible with these
restrictions if we also have $\int_{-\infty}^{+\infty} d\sigma^r\,
r_{\bar a}(\tau ,\vec \sigma ) = 0$.

The integral equations  for the lapse and shift functions yield
the following solutions [the signature of the 4-metric is
$\epsilon (+---)$ with $\epsilon =\pm 1$]  $N(\tau ,\vec \sigma )=
-\epsilon +n(\tau ,\vec \sigma )=-\epsilon + O(r^2_{\bar a})$,
$N_r(\tau ,\vec \sigma )= n_r(\tau ,\vec \sigma ) =-\epsilon \,
{}^4{\hat g}_{\tau r}(\tau ,\vec \sigma )$ and
\begin{widetext}
\bea
 n_r(\tau ,\vec \sigma )
 &=& {{\partial}\over {\partial \sigma^r}}\,
      \Big(  {{2\sqrt{3}\pi G}\over {c^3}}
 \sum_{\bar av} \gamma_{\bar av}\nonumber
 \Big[ \!\! \sum_{ua,u\not= a}
 [1\!-\!2(\delta_{uv}\!+\!\delta_{av})]\,
 \int^{\sigma^u}_{-\infty}  \!\!\!\! d\sigma_1^u
 \int^{\sigma^a}_{-\infty}  \!\!\!\! d\sigma_1^a
 \int^{\infty}_{\sigma_1^u} \!\!\!\! d\sigma_2^u
 \int^{\infty}_{\sigma_1^a} \!\!\!\! d\sigma_2^a
 {{\partial^2 \pi_{\bar a}(\tau ,\sigma_2^u \sigma_2^a
 \sigma_2^{k\not= u,a})}\over {(\partial \sigma_2^v)^2}}
 {|}_{\sigma_2^k=\sigma^k}-
\nonumber \\
 &-& 2\sum_{u\not= r} [1-2(\delta_{uv}+\delta_{rv})]\,
 \int^{\sigma^r}_{-\infty}  \!\!\!\!d\sigma_1^r
 \int^{\sigma^u}_{-\infty}  \!\!\!\!d\sigma_1^u
 \int^{\infty}_{\sigma_1^r} \!\!\!\!d\sigma_2^r
 \int^{\infty}_{\sigma_1^u} \!\!\!\!d\sigma_2^u
 {{\partial^2 \pi_{\bar a}
 (\tau ,\sigma_2^r \sigma_2^u \sigma_2^{k\not= r,u})}\over
 {(\partial \sigma_2^v)^2}} {|}_{\sigma_2^k=\sigma^k}\Big] \Big)+
  O(r^2_{\bar a}).
 \label{3}
\eea
\end{widetext}
After the solution of all the constraints (super-hamiltonian one
included), the {\it 4-metric} of our linearized space-time, in
$\Sigma^{(WSW)}_{\tau}$-adapted coordinates, in our 3-orthogonal
gauge,  becomes (the form of a perturbation of the Minkowski
metric in Cartesian coordinates is used {\it only to visualize}
the deviations from special relativity)  ${}^4{\hat g}_{AB}(\tau
,\vec \sigma ) = {}^4\eta_{AB} + {}^4h_{AB}(\tau ,\vec \sigma )$.
We have ${}^4h_{\tau\tau}(\tau ,\vec \sigma )= 0 + O(r^2_{\bar
a})$, ${}^4h_{\tau r}(\tau ,\vec \sigma )  = -\epsilon n_r(\tau
,\vec \sigma )$ and
\begin{widetext}
\bea
 {}^4h_{rs}(\tau ,\vec \sigma ) &=&
 -{{2\epsilon}\over {\sqrt{3}}} \sum_{\bar a} \Big[ \gamma_{\bar ar}
 r_{\bar a}(\tau ,\vec \sigma ) -
 {1\over 2} \sum_u\gamma_{\bar au} \int d^3\sigma_1 {{\partial^2_{1u}r_{\bar a}(\tau
 ,{\vec \sigma}_1)}\over {4\pi |\vec \sigma -{\vec \sigma}_1|}} \Big]
 \delta_{rs} + O(r^2_{\bar a}).
\label{4}
\eea
As said the Hamiltonian is the weak ADM energy. After the solution
of all the constraints its quadratic part in the DO's is [see
Eq.(2.15) of Ref.\cite{2} for the xpression of the kernel ${\cal
T}^{(o)u}_{(a)r}(\vec \sigma ,{\vec \sigma}_1 $]
 \bea
 {\hat E}_{ADM}
 &=& {{12\pi G}\over {c^3}} \int d^3\sigma \sum_{\bar a}
  \bigg[
  \pi^2_{\bar a}(\tau ,\vec \sigma ) +
  \sum_{\bar a\bar b} \sum_{rs} \gamma_{\bar ar}\gamma_{\bar bs}
 \int d^3\sigma_1 d^3\sigma_2 \,  \sum_u {\cal T}^{(o)u}_{(a)r}(\vec \sigma
 ,{\vec \sigma}_1) {\cal T}^{(o)u}_{(a)s}(\vec \sigma ,{\vec
 \sigma}_2)\,  \pi_{\bar a}(\tau ,{\vec \sigma}_1)\,
   \pi_{\bar b}(\tau ,{\vec \sigma}_2)
  \bigg]
  -\nonumber \\
 &-& {{c^3}\over {16\pi G}} \sum_r \int d^3\sigma
 \Big[ {1\over 6} \Big( \sum_{\bar au} \gamma_{\bar au}
 {{\partial}\over {\partial \sigma^r}} \int d^3\sigma_1
 {{\partial^2_{1u}r_{\bar a}(\tau ,{\vec \sigma}_1)}\over
 {4\pi |\vec \sigma -{\vec \sigma}_1|}} \Big)^2 -
{1\over 3}\sum_{\bar a} \Big( \partial_rr_{\bar a}(\tau ,\vec \sigma )\Big)^2+
 {2\over 3} \Big( \sum_{\bar a}\gamma_{\bar ar} \partial_rr_{\bar a}(\tau ,\vec \sigma )
 \Big)^2 -
\nonumber \\ &-&
{1\over 3} \sum_{\bar a\bar b} \sum_u \gamma_{\bar ar} \partial_rr_{\bar a}(\tau
 ,\vec \sigma ) \gamma_{\bar bu} {{\partial}\over {\partial \sigma^r}}
 \int d^3\sigma_1 {{\partial^2_{1u}r_{\bar b}(\tau ,{\vec \sigma}_1)}\over
 {4\pi |\vec \sigma -{\vec \sigma}_1|}} \Big] + O(r^3_{\bar a}).
 \label{5}
\eea
\end{widetext}
The rest-frame condition, i.e. the vanishing of the weak ADM
3-momentum, is ${\hat P}^r_{ADM} = - \int d^3\sigma\, \sum_{\bar
c}\, \pi_{\bar c}(\tau ,\vec \sigma )\, \partial_r\, r_{\bar
c}(\tau ,\vec \sigma ) \approx 0$. Since we are in an instant form
of the dynamics both the ADM 3-momentum and angular momentum
(${\hat J}^{rs}_{ADM} =\int d^3\sigma\, \sum_{\bar c}\,
 \pi_{\bar c}(\tau ,\vec \sigma )\, (\sigma^r\, \partial_s -
 \sigma^s\, \partial_r)\, r_{\bar c}(\tau ,\vec \sigma )$) have the
same form as in a free theory. Instead the ADM energy and boosts
have a complicated form. Notwithstanding this fact, the study of
the Hamilton equations for the DO's imply \cite{2} that, even if
we are not in a harmonic gauge, the $r_{\bar a}(\tau ,\vec \sigma
)$ satisfy the wave equation
\beq
 \Box\, r_{\bar a}(\tau ,\vec \sigma )\, \cir\, 0.
 \label{6}
\eeq
A {\it class of solutions} of the Hamilton equations
vanishing correctly at spatial infinity, satisfing
the rest frame condition ${\hat P}^r_{ADM} = 0$ and every
restriction, is
\begin{eqnarray}
\label{7}
&&\!\!\!\!
r_{\bar a}(\tau ,\vec \sigma ) = \\
&&= C_{\bar a} \int
      \!\!\! {{d^3k}\over {(2\pi)^3}}
      {{(k_1 k_2 k_3)^2 e^{- {\vec k}^2} }\over {|\vec k|}}
 \!\Big[ e^{-i (|\vec k| \tau \!-\! \vec k \cdot \vec \sigma )}
     \! +  \! e^{ i (|\vec k| \tau \!-\! \vec k \cdot \vec \sigma )} \!\Big]
\nonumber \\
&&= - {{4\, C_{\bar a}}\over {(2\pi )^2}}\,
 {{\partial^6} \over
 {\partial^2 \sigma_1}
 {\partial^2 \sigma_2}
 {\partial^2 \sigma_3}}\,
 {{(1 + |\vec \sigma |^2 - \tau^2)}\over
  {[\! 1 \!+\! (|\vec\sigma | \!+\! \tau )^2]\,
   [\! 1 \!+\! (|\vec\sigma | \!-\! \tau )^2]}},
\nonumber \\
&&\!\!\!\!
\pi_{\bar a}(\tau ,\vec \sigma ) = -i\,  \int {{d^3k}\over {(2\pi
 )^3}}\, {{(k_1 k_2 k_3)^2}\over {|\vec k|}}\, e^{- {\vec
 k}^2}\, |\vec k| \\
&& \, \sum_{\bar b}\, A^{-1}_{\bar a\bar b}(\vec k)\,
 C_{\bar b}\,  \Big[ e^{-i (|\vec
 k|\, \tau - \vec k \cdot \vec \sigma )} -  e^{i (|\vec
 k|\, \tau - \vec k \cdot \vec \sigma )} \Big].
\nonumber
\end{eqnarray}
\noindent It describes standing waves and the two constants
$C_{\bar 1}$, $C_{\bar 2}$ have to be expressed in terms of the
two boundary constants $M = {\hat E}_{ADM}$, $S = |{\hat {\vec
J}}_{ADM}|$, defining the mass and spin of the post-Minkowskian
space-time. {\it In absence of matter, the rest-frame
condition destroys the transversality property of the TT harmonic
gauge plane waves}.

\begin{figure}
\vspace{1mm} \centerline{\includegraphics[width=8cm]{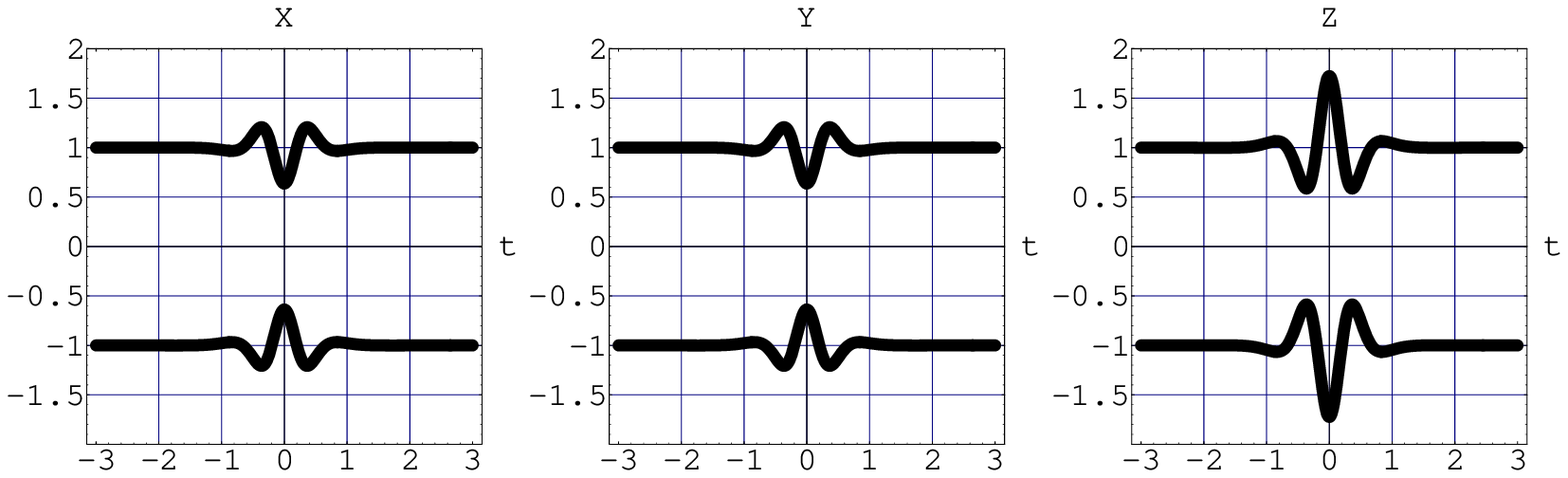}}
\vspace{2mm} \centerline{\includegraphics[width=8cm]{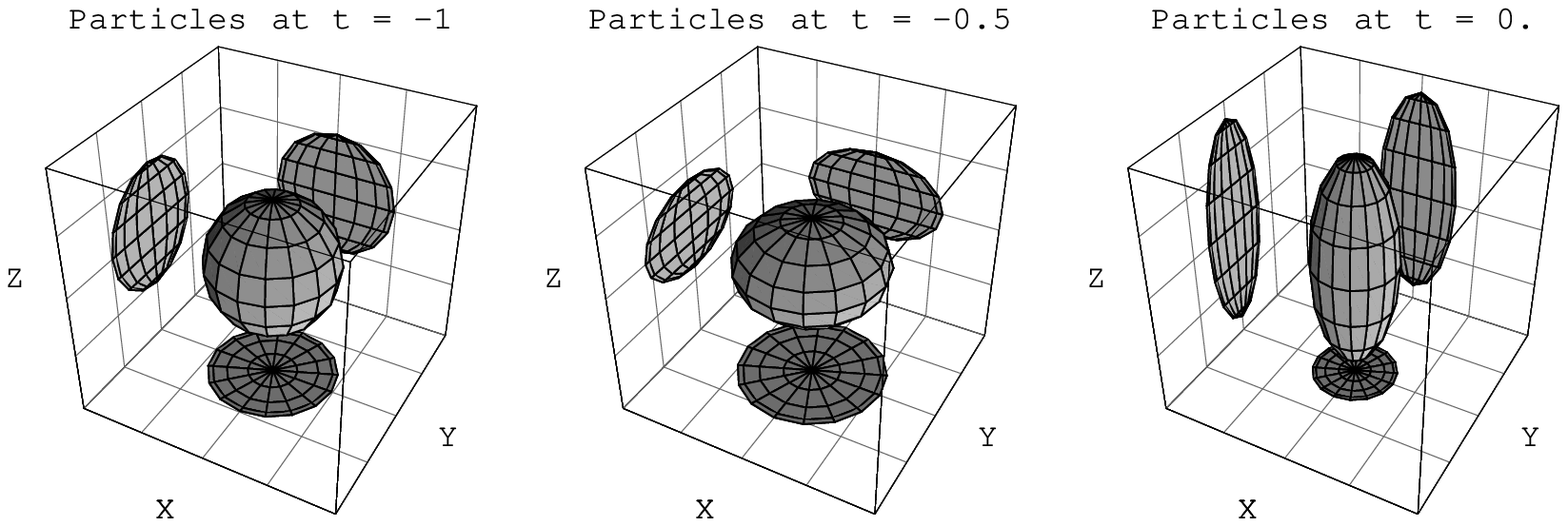}}
\vspace{2mm} \caption{Deformation of a sphere of particle at rest
induced by the passage of the gravitational wave packet for
$C_{\bar 1} \not= 0$.} \label{fig:R1}
\end{figure}

\begin{figure}
\vspace{1mm} \centerline{\includegraphics[width=8cm]{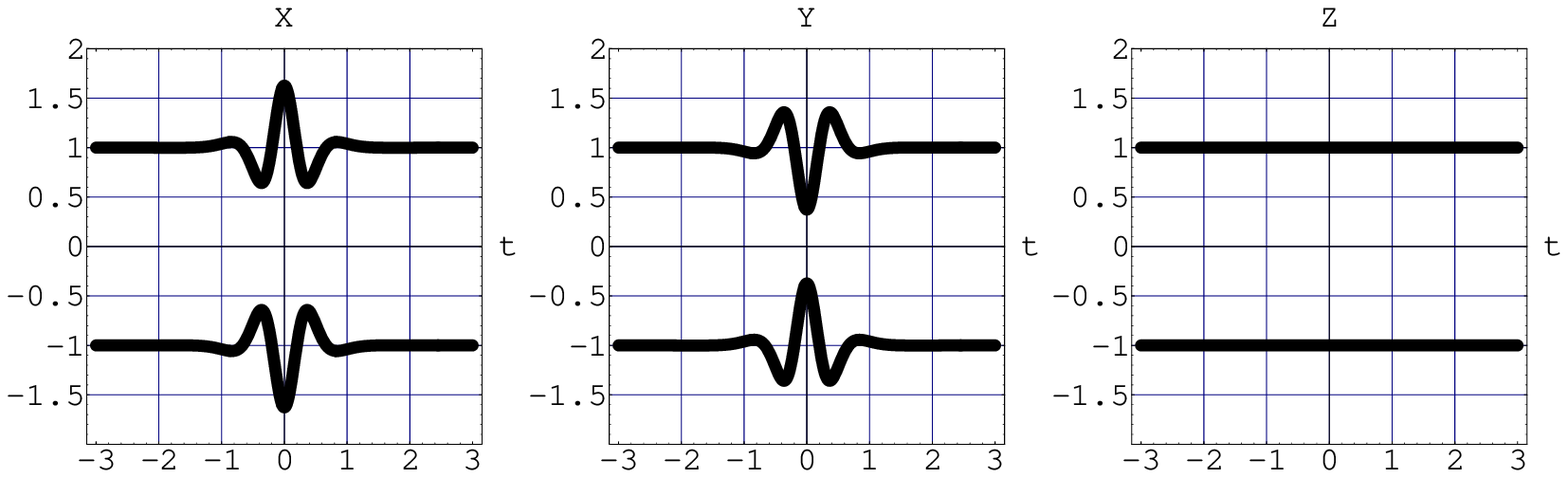}}
\vspace{2mm} \centerline{\includegraphics[width=8cm]{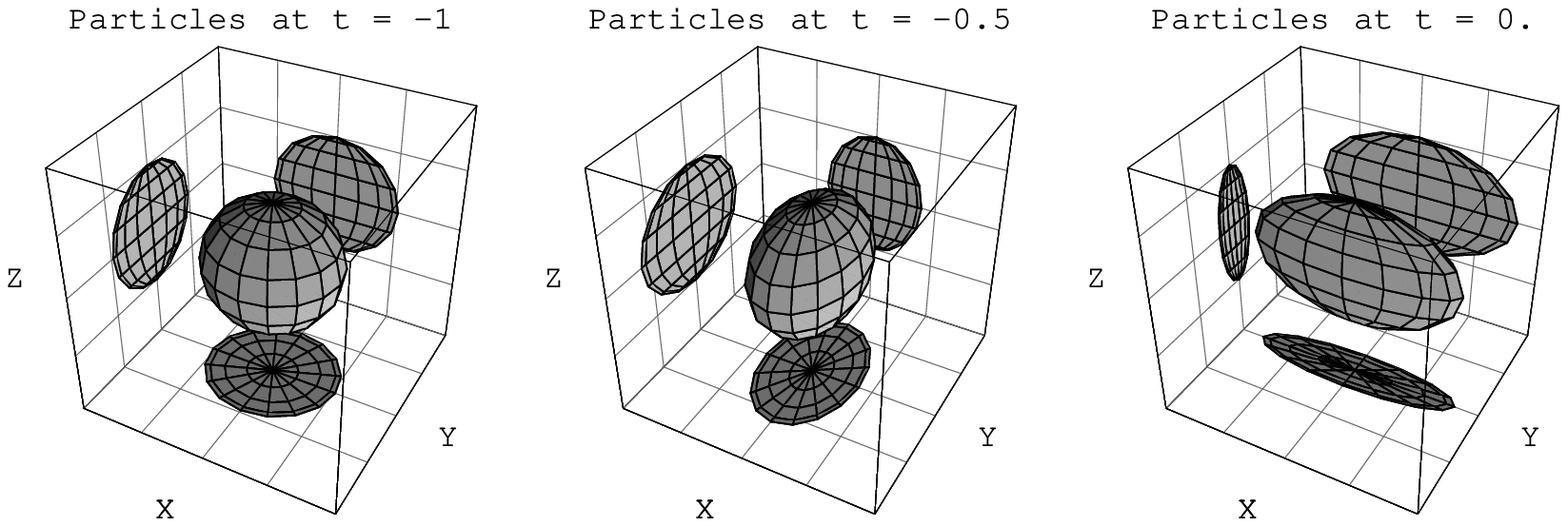}}
\vspace{2mm} \caption{Deformation of a sphere of particle at rest
induced by the passage of the gravitational wave packet for
$C_{\bar 2} \not= 0$.} \label{fig:R2}
\end{figure}

In Ref.\cite{2}, after a study of both the geodesic equation and
the geodesic deviation equation in our gauge, we solve the latter
equation numerically for a sphere of test particles at rest around
the origin of the 3-coordinates on a WSW hyper-surface for the
previous solution. We obtain the two 3-dimensional deformation
patterns replacing the usual 2-dimensional ones for the
polarization in the TT harmonic gauge:
i) in figure \ref{fig:R1} there is the deformation pattern for the
case $C_{\bar 1} \not= 0$, $C_{\bar 2} = 0$, namely for $r_{\bar
1}(\tau ,\vec \sigma ) \not= 0$, $r_{\bar 2}(\tau ,\vec \sigma ) =
0$;
ii) in figure \ref{fig:R2} that for
the case $C_{\bar 1} = 0$, $C_{\bar 2} \not= 0$, namely for
$r_{\bar 1}(\tau ,\vec \sigma ) = 0$, $r_{\bar 2}(\tau ,\vec
\sigma ) \not= 0$.
In the two figures are reported the snapshots at three
different times ($t = -1, -0.5, 0$) of the sphere of particles
originally at rest (bottom) and the time evolution (from $t=-3$ to
$t=3$) of the six particles at the intersection of the three axes
and the sphere of particle (top), whose initial 3-coordinates are
$(1,0,0)$ and $(-1,0,0)$ on the $x$-axis, $(0,1,0)$ and $(0,-1,0)$
on the $y$-axis,  $(0,0,1)$ and $(0,0,-1)$ on the $z$-axis,
respectively. Only the $i$ coordinates are reported for the two
particles on the $i$ axis, since they remain on it.

Let us remark that till now we have a treatment of the generation
of gravitational waves from a compact localized source of size $R$
and mean internal velocity $v$ only \cite{11} for {\it nearly
Newtonian slow motion} sources for which $v << c$,
${{\lambda}\over {2\pi}} >> R$: outgoing gravitational waves are
observed in the {\it radiation zone} (far field, $r >>
{{\lambda}\over {2\pi}}$), while deep in the {\it near zone} ($R <
r << {{\lambda}\over {2\pi}}$), for example $r \leq 1000 \,
{{\lambda}\over {2\pi}}$, vacuum Newtonian gravitation theory is
valid. On the contrary with our approach in suitable 4-coordinates
we are going to obtain a {\it weak field approximation but with
fast relativistic motion in the source} subject to the restriction
that the total invariant mass and the mass currents are compatible
with the weak field requirement. This is enough to get
relativistic results conceptually equivalent to the re-summation
of the post-Newtonian expansion. Therefore the results about
background-independent gravitational waves in post-Minkowskian
space-times   {\it  open the possibility, after the introduction
of matter, to study the emission of gravitational waves from
relativistic sources without any kind of post-Newtonian
approximation}. For instance this is the case for the relativistic
motion (but still in the weak field regime) of the binaries before
the beginning of the final inspiral phase: it is known that in
this phase the post-Newtonian approximation does not work and
that, till now, only numerical gravity may help. In a future paper
we will add a relativistic perfect fluid, described by suitable
Lagrangian \cite{12} or Eulerian \cite{13} variables, to tetrad
gravity, we will define a Hamiltonian linearization of the system
in our completely fixed 3-orthogonal gauge, we will find the
Hamilton equations for the DO's both of the gravitational field
and of the fluid, we will find the relativistic version of the
Newton and gravito-magnetic action-at-a-distance potentials and of
the generalized tidal effects acting inside the fluid and finally,
by using a multipolar expansion, we will find the relativistic
counterpart of the post-Newtonian quadrupole emission formula.

Moreover we will have to explore whether our Hamiltonian approach
is suitable for doing  Hamiltonian numerical gravity on the full
non-linearized theory.


\end{document}